**Fermi surface origin of the low-temperature magnetoresistance anomaly**

Yejun Feng[1,*], Yishu Wang[2,3], T. F. Rosenbaum[4], P. B. Littlewood[5], Hua Chen[6]

[1]Okinawa Institute of Science and Technology Graduate University, Onna, Okinawa 904-0495, Japan
[2]Department of Materials Science and Engineering, University of Tennessee, Knoxville, Tennessee 37996, USA
[3]Department of Physics and Astronomy, University of Tennessee, Knoxville, Tennessee 37996, USA
[4]Division of Physics, Mathematics, and Astronomy, California Institute of Technology, Pasadena, CA 91125, USA
[5]The James Franck Institute and Department of Physics, The University of Chicago, Illinois 60637, USA
[6]Department of Physics, Colorado State University, Fort Collins, Colorado 80523, USA
*Corresponding author. Email: yejun@oist.jp

**Abstract:**

A magnetoresistance (MR) anomaly at low temperatures has been observed in a variety of systems, ranging from low-dimensional chalcogenides to spin and charge density wave (SDW/CDW) metals and, most recently, topological semimetals. In some systems parabolic magnetoresistance can rise to hundreds of thousands of times its low-temperature, zero-field value. While the origin of such a dramatic effect remains unresolved, these systems are often low-carrier-density compensated metals, and the physics is expected to be quasi-classical. Here we demonstrate that this MR anomaly in temperature also exists in high conductivity good metals with large Fermi surfaces, namely Cr, Mo, and W, for both linear and quadratic field-dependent regimes with their non-saturation attributed to open orbit and electron-hole compensation, respectively. We provide evidence that quantum transport across sharp Fermi surface arcs, but not necessarily the full cyclotron orbit, governs this low-temperature MR anomaly. In Cr, extremely sharp curvatures are induced by superposed lattice and SDW band structures. One observes an overlay of the temperature dependence of three phenomena: namely, MR at a constant high field, linear MR in the low-field limit, and Shubnikov-de Haas (SdH) oscillations of the lightest orbit. In Mo, the temperature dependence of low-$T$ MR anomaly extends beyond those of its SdH oscillations but disappears at temperatures where Kohler's scaling reemerges. In the low-temperature and high-field limit, large magnetoresistance from carriers circling quantum orbits is the three-dimensional analogy to the zero-conductance state of carrier localization in the integer quantum Hall effect, especially with regard to the adverse effect of disorder.

Introduction

One major characteristic of metals is their thermal behavior. As temperature is reduced, a metal often demonstrates improved electrical and thermal conductivities; the opposite is typical for insulators. However, under a large magnetic field, this conclusion is challenged. In recent decades, many reports have emerged across semi-metallic and intermetallic conductors demonstrating a very large MR ratio (over $10^4$ at ~10T) [1]. Graphite, bismuth [2], $WTe_2$ [3], $Cd_3As_2$ [4], LaSb, LaBi [5], and ZrSiS [6] all demonstrate anomalously large, positive, and quadratic magnetoresistance in the zero-temperature limit. While the extremely large MR is often at the center of attention, the MR under a constant magnetic field follows a non-monotonic temperature dependence and strikingly appears as an anomaly at low temperature [7]. This rise of magnetoresistance $\rho(T, H=\text{const.})$ as $T$ goes to zero has been interpreted as metal-insulator transition [2] or different types of phase transitions [5], and even posited to be linked to topological mechanisms [1, 4]. But its fundamental origin and the question of universality remain unsettled [1]. For these linked phenomena we use the term "the MR anomaly" in this paper, noting that it is unrelated to zero-field phenomena with the moniker "anomalous Hall effects".

Many semimetals and intermetallic compounds of interest have a large MR with $H^2$ field dependence; some but not all may be topologically non-trivial [1]. This leads to an explanation for semimetals based on an electron-hole compensation mechanism [1], where carriers of the hole and electron types are of the same density. While electron-hole compensation can explain the $H^2$ functional form, the large magnitude of MR in semimetals such as Bi is attributed to its extremely small carrier density $n$ that leads to a large $\omega_c\tau$ [8], where $\omega_c$ is the cyclotron frequency and $\tau$ the carrier relaxation time. Furthermore, it is well known that disorder has an adverse effect on the size of MR, and the MR ratio is strongly correlated to the *RRR* value [1, 3, 7]. It is then argued that disorder can vary the compensation condition in semimetals to reduce the MR magnitude in nominally identical systems [1].

However, extremely large magnetoresistances at low temperature have been reported in highly conductive elemental metals such as Cr [9], Mo, and W [10, 11]. Comparing the MR ratio $\rho(T=4K, H)/\rho(T=4K, H=0)$ and residual resistivity ratio (*RRR*) $\rho(T=300K, H=0)/\rho(T=4K, H=0)$, a MR anomaly is expected to exist in those systems, although the anomalous temperature dependence at a fixed field was not revealed in Cr until recently [7] and in Mo and W in this study. For good metals ($\rho_0$~0.1 μΩ cm) with an overall carrier density in the range of $10^{23}/cm^3$, the global $\omega_c\tau$ is very small. So typically, no large MR is expected [8]. Furthermore, compensation in highly conductive metals [12] is satisfied to a high precision [11] and the large carrier density makes the compensation condition unlikely to be altered by disorder.

In both intermetallic/semiconductive materials [1, 3, 4, 13] and SDW/CDW systems [7], the low-$T$ magnetoresistance anomaly is a universal phenomenon for MR both large and small. Furthermore, the large MR effect is always most prominent at low $T$. So understanding the low-$T$ MR anomaly from the temperature perspective can lead to clarity about the underlying mechanism. Instead of affecting the compensation, we argue that disorder's negative effect and the low-$T$ prominence of the MR anomaly arise from its quantum nature, with connections to Landau-level cyclotron physics [8, 14] and features of the Fermi surface. For partially gapped SDW/CDW metals, a large and linear MR at low temperature, as well as the more usual quadratic MR, can result from charge carriers transiting extremely sharp corners at the Fermi surface, but not necessarily completing the full orbit around a closed surface [7, 8].

Here we survey detailed origins of the MR in the cubic metals Cr and Mo over a wide temperature range. Both systems support unsaturated MR because of either open orbit or electron-hole compensation. In Cr, sharp curvature on a small cyclotron orbit is identified through the overlay of three different temperature dependences: that of the MR at 14 T, the linear MR below 0.3 T, and SdH oscillations at ~5 T. In Mo, the MR anomaly has a characteristic temperature bound by two scenarios. First, it exists beyond the temperature all SdH oscillations disappear, indicating carriers do not need to complete the full cyclotron motion. Second, Kohler's scaling, indicative of a momentum-independent relaxation time, fails at low temperature but reemerges at a temperature higher than the MR anomaly's characteristic temperature. Taken together, the evidence points to a MR anomaly that is bound to sharp local curvature of the Fermi surface. The competition between the inverse of the sharp corner's curvature and the carrier's relaxation length explains the low-$T$ MR anomaly while both electron-hole compensation and the open orbit remain unchanged over the pertinent temperature ranges in both Cr and Mo. In the high-field and low-$T$ limit, this mechanism's quantum nature and sensitivity to disorder suggest the extraordinarily high value of MR is an analogy to carrier localization of the integer quantum Hall effect in the two-dimensional electron gas.

**RESULTS**

**Low-$T$ MR anomaly and disorder's adverse effect**

Transverse magnetoresistance $\rho_{xx}(H_z)$ of the cubic elemental metals Cr, Mo, and W (Methods) are presented in Fig. 1, revealing the non-monotonic temperature dependence of the low-$T$ MR anomaly at 14 T. All three systems are group-VI elements of identical $d$-electron configurations and possess isomorphic crystal structures of a simple body-centered cubic Bravais lattice. Their paramagnetic Fermi surfaces are well understood as composed of closed forms [15] and differ only in the degree of spin-orbit coupling. Our Mo and W single crystal specimens demonstrate a low-temperature MR rising by a factor of $\Delta\rho/\rho_0 = \rho(1.65\text{K}, 14\text{T})/\rho_0(1.65\text{K}, 0\text{T}) - 1 \approx 14{,}000$ and 145,000, respectively (Fig. 1), which is comparable to many semi-metallic systems [1]; WTe$_2$ has a MR ratio $\Delta\rho/\rho_0$ varying from 50 to 17,000 at 9 Tesla [3]. Here Mo and W exemplify the fact that the low-$T$ MR anomaly is not a phenomenon limited to intermetallic compounds, semimetals, density waves, and low-dimensional systems [1-3], but is also present in highly conductive three-dimensional simple metals.

In our single crystal Mo specimen, $RRR = \rho(300\text{K}, 0)/\rho(2\text{K}, 0) = 2900$, and the MR ratio reaches beyond $10^4$ (Fig. 1), far exceeding the MR ratio of only ~40 in a polycrystalline Mo specimen of $RRR$=160 (Fig. 1). As the MR effects only have an anisotropy factor about two in Mo single crystals [10], the difference between single and polycrystalline samples is not going to account for the significant reduction of MR in samples with lower $RRR$. Similar results are also observed in W. For a specimen of $RRR$~10,000, the MR ratio exceeds 145,000, while for a polycrystal sample of $RRR$~12 the low-$T$ MR anomaly is barely observable (Fig. 1). Our W single crystal specimen might have a non-trivial level of amorphous W, which can be superconducting at low $T$ [16]. The amount is not large enough to cause the whole sample to superconduct, but it does create difficulties in precisely determining $\rho_0$ in MR measurements at low $T$ (Methods). From here on, we focus on Cr and Mo. $\rho(T)$ of our single crystal Cr and Mo specimens exhibits a $T^3$ power law dependence in the low $T$ limit (Fig. 1), indicating a disorder-dominated instead of a phonon-dominated scattering mechanism.

Several studies in the literature [5, 17] have sought to explain the meaning of the temperature position of the $\rho(T, H=\text{const.})$ minimum in the MR anomaly. The simplest explanation is that the minimum in $\rho(T, H=\text{const.})$ is simply the intersection of two behaviors: namely, a large low-temperature MR that decreases with rising $T$ and a thermally (phonon scattering) induced $\rho(T, 0)$ that increases with rising $T$. Nevertheless, the contrast between low temperature and thermal effects is suggestive of the quantum nature of the low-$T$ MR. In the discussion below, we explore the characteristic temperature scales of the non-monotonic $\rho(T, H=\text{const.})$ in both Cr and Mo.

**Sharp curvatures on a small orbit of SDW Cr**

Cr develops a partially gapped Fermi surface due to a spin density wave (SDW) below 311.5 K. As a universal phenomenon of DW states [7], Cr has a linear MR at both low field and low temperature (Fig. 2). With contributions from carriers of different parts of the Fermi surface, $\rho(T,H)$ can be separated into a sum of linear and quadratic terms as $\Delta\rho(T,H) = \Delta\rho_{lin} + \Delta\rho_{quad} = A(T)|H| + B(T)H^2$. Here the summation form of $\Delta\rho$ is justified as the MR $\Delta\rho(T,H)$ is analyzed in the low field region and only reaches a fraction of $\rho_0 = \rho(T, H=0)$. In Fig. 2b, both linear and quadratic contributions of $\Delta\rho$ are plotted against $\rho_0(T)$ and the linear component monotonically reduces to zero with increasing $\rho_0(T)$.

As carriers on each Landau tube contribute simultaneously to both the MR and SdH oscillations, an individual band's contribution to the MR anomaly can be marked by its SdH frequency [14]. In Cr, the incommensurate SDW wave vector **Q** creates gapped, fragmented Fermi surfaces to form very small quantum orbits [9, 18]. Quantum oscillations in Cr were recently studied in detail, but their temperature dependences were not explored fully [18]. Here, representative curves of $\Delta\rho_{\text{SdH}}(H)/\rho(H)$ at various $T$ are plotted versus $1/H$ in Fig. 2. While all dHvA oscillations disappear at $T \sim 25$ K, SdH oscillations of a quantum interference nature, most noticeably the 36 T orbit, persist to temperature beyond 90 K [18]. In Fig. 2, the SdH amplitude $\Delta\rho_{SdH}/\rho$ of the 36T oscillation is extracted and its temperature dependence largely follow the Lifshitz-Kosevich formula for $T \geq 6$ K; its value at 1.65 K (open circle in Fig. 2d) is about 10% smaller than the 6 K value due to contributions to $\rho(H)$ by other SdH orbitals at low $T$ [19].

Our galvanomagnetic measurements of the single-Q-domain Cr crystal have generated temperature dependences of three phenomena: the MR anomaly at 14 T (Fig. 1) $\Delta\rho(T, H=14\text{T})/\rho(T, H=0)$, the linear MR in the low field limit (Fig. 2b) $\Delta\rho_{lin}(T, H=0.3\text{T})/\rho(T, H=0)$, and the 36 T SdH oscillation (Fig. 2c) amplitude $\Delta\rho_{SdH}(T,H)/\rho(T,H)$. All three temperature dependences are similar in shape with a specific and non-trivial temperature scale of ~40 K at the half maximum (Fig. 2d). This connection between all three galvanomagnetic behaviors provides direct evidence that transport on the 36 T orbit controls the low-$T$ MR anomaly.

For MR of Cr measured at the base temperature (Fig. 2e), there is a region near $H$=0 that can be fit fully to a parabolic form as $\Delta\rho(T,H) = BH^2$, without the symmetric linear component. The transition field $H_c$ from parabolic to linear MR behavior sets the condition of $\omega\tau \sim 2\pi$, with $\omega$ being the cyclotron frequency at the most dominant local curvature. Equivalently, the carriers' free-path length $l$ would be comparable to the real space orbital radius $r$ as $l/r \sim \omega\tau$ [8, 18]. For our Cr sample with $RRR_{350K} \sim 70$ and $\rho_0 \sim 206$ nΩ cm, $H_c$ is ~135 Oe at 1.65 K (Fig. 2e). Following the Onsager-Lifshitz relationship $r_k = (eH/ch)r$, we have a plausible $l \sim r \sim 220$ Å, if $r_k \sim$ 3 $10^{-5}$ Å$^{-1}$. As $l$ reduces with rising temperature, a diminishing $r$ follows an increasing $H_c$, which makes the MR gradually transition from a linear to a parabolic form over a fixed field range (Fig. 2a).

While the presence of linear MR down to $H_c$ of 10-100 Oe is common in SDW/CDW systems as a signature of extremely sharp local curvatures [7], our observed $H_c$ (Fig. 2e, Ref. [7]) are very different from theoretical estimations of ~2T in the literature [20, 21]. In DW systems, extremely sharp curvatures generically exist at the edges of partially opened gaps, where carriers move between different pieces (segments) of Fermi surfaces determined by two superposed reciprocal lattices [18, 20, 21]. The curvature at the edge of the DW induced energy gap is set by the inverse of the spatial coherence lengths $\xi$ of both the lattice and DW, which determines the resolution of the reciprocal space. For Cr as a high-quality single crystal, $\xi$ can be ~ 1-10 μm for a single SDW domain, setting the sharpest reciprocal space curvature, $r_k \sim 1/\xi \sim 10^{-4}$-$10^{-5}$ Å$^{-1}$. The 36 T SdH orbit in Cr has an average radius ~0.003 Å$^{-1}$, so the sharpest curvature with a radius $r_k \sim 3$ $10^{-5}$ Å$^{-1}$ would become the most significant contributor to the MR anomaly.

While our Cr specimen has a modest $RRR_{350K} \sim 70$ and a MR ratio ~50 at 1.65 K, high quality Cr specimens ($RRR \sim 750$) in the literature have produced large MR ratios of 1000 to 1800 at 8.5 T [9]. As the MR has a power law exponent close to 1 (about 1.3 in Ref. [9] which is consistent with the power law behavior of MR in our sample; see below), the sharp corner mechanism certainly can produce very large MR in highly conductive metals.

**SdH oscillations and sharp arcs in Mo**

Contributors to the MR anomaly in Mo also can be revealed by SdH oscillations. Mo's Fermi surface is composed of only closed forms and there is no sub-500T dHvA orbit [22]. We measure SdH oscillations of Mo under the geometry of field **H** parallel to one cubic axis and current **I** parallel to another (Fig. 3) and observe all dHvA frequencies under this geometry [15, 22-23], in addition to many higher harmonics (Fig. 3a). At 9 K, nearly every SdH oscillation, except three, have vanished, while the MR anomalies $\Delta\rho(14T)/\rho(0)$ still have ~80% of the strength in the low-$T$ limit (Fig. 3).

Two SdH orbits of 518 T and 1688 T are the only surviving quantum oscillations beyond $T \sim 12$ K (Fig. 3d). The 518 T oscillation has the largest SdH amplitude $\Delta\rho_{SdH}/\rho$ at 1.65 K and has a monotonic temperature dependence that follows the Lifshitz-Kosevich form very well from 6 K on (Fig. 3d). The 1688 T oscillation is comparatively weak at 1.65 K, but it is the only SdH oscillation to survive beyond 18 K, present up to 28 K (Fig. 3d). However, this orbit is absent in dHvA measurements, and its frequency instead suggests a combination of two dHvA orbits, 518T and 1170T. As it exceeds every other dHvA orbit's temperature dependence, the 1688 T oscillation is consistent with quantum interference [19]. The 518 and 1170 T oscillations are attributed to two orbits in proximity, named electron lens and neck, respectively [15, 22-23]. They are not regarded as coplanar in the literature [15] but would be so if the combination is true; the electron lens would then be a prolate ellipsoid, instead of an oblate ellipsoid [15].

In both Cr and Mo, we notice that every SdH orbit (Figs. 2d, 3e) follows a Lifshitz-Kosevich form that is no broader than the temperature dependence of the MR anomaly $\Delta\rho(14T)/\rho(0)$. This is similarly observed in other systems in the literature, although many have used dHvA techniques [1, 5-6]. SdH studies reveal quantum orbits of both the Landau level type and quantum interference nature [18], while dHvA measurements would not be sensitive to transport features such as the 36 T and 1688 T orbits in Cr and Mo respectively. Despite being a more challenging measurement and not favored in the literature [1], SdH techniques have advantages over dHvA techniques in exploring the low-$T$ MR anomaly. Comparisons of the temperature dependences of SdH oscillations and MR confirm that the carriers do not need to complete the full cyclotron motion, but only to transit a sharp arc in order to generate a large MR [7-8].

The 518 T orbit is the common feature of the 518 and 1688 T SdH orbits in Mo, so the sharp curvature is expected to be on the 518 T orbit, but not at the magnetic breakdown point to the 1170 T orbit. The sharpest curvature is again estimated using the Onsager-Lifshitz relationship. In our Mo specimen, the critical field $H_c$ for the limit of parabolic MR is ~420 Oe (Fig. 3). To estimate $r$ (or $l$), our Mo specimen has a 100× better residual resistivity $\rho_0$ than that of Cr (1.9 vs. 206 nΩ cm) at 1.65 K. From the free electron picture [8], $l$ and $\rho$ are related by $\rho l \sim n^{-2/3}$, where $n$ is the carrier density, and because of the SDW gap, $n$ in Cr is about half that of Mo. The very low value of $\rho_0$ in Mo suggests $l \sim 10,000$ Å, leading to $r_k \sim 0.004$ Å$^{-1}$ for the sharpest curvature under the **H** || (1,0,0) geometry. In comparison, the 518 T orbit has an average radius of ~0.125 Å$^{-1}$. In the angular study of MR in Mo (Fig. 2 of Ref. [10]), the MR is the smallest with **H** || (1,0,0), exhibiting a power law exponent of 1.8 in comparison to other directions such as those 30-degrees off the (1,0,0) axis, which have a power law exponent of 2.0 [10]. A reduced power law exponent suggests the 518 T orbit has a curvature that is sharper than the average.

**Kohler's scaling**

We notice that SdH oscillations of all orbits set a lower temperature boundary for the origin of the MR anomaly. Transport phenomena of a quantum nature are different from scattering-based semiclassical descriptions, as SdH oscillations would only depend on the field $H$ but not $\rho_0$ [8]. It is thus instructional to examine Kohler's scaling, based on semiclassical Boltzmann transport theory, in which $\Delta\rho/\rho_0$ at various temperatures collapse onto a single master curve $f(H/\rho_0)$. Kohler's plot holds when there is only one momentum-independent $\tau$ at each temperature for carriers of all Fermi surfaces. It exists in a region where quantum effects disappear but phonon scattering does not yet dominate [24]. A failure of Kohler's scaling was observed in many systems with low-$T$ MR anomaly such as NbSe$_2$ [13], Cu$_{2-x}$Te [25], and TaP [26].

In the literature, $\Delta\rho(H)/\rho_0$ is often characterized by a power law functional form of $\Delta\rho/\rho_0 \sim H^\alpha$. In Fig. 4, $\Delta\rho/\rho_0$ vs. $H/\rho_0$ is plotted for various temperatures, allowing simultaneous examinations of both the power law form and Kohler's scaling. For single-Q Cr, the power exponent $\alpha(T)$ varies from 1.3 to 1.5 between 1.65 and 180 K; $\alpha$ is about 1.8 for Mo under our measurement geometry [10]. We also make a "residual" plot of $\frac{\Delta\rho}{\rho_0}/(\frac{H}{\rho_0})^{\alpha_0}$ vs. $H/\rho_0$, with $\alpha_0$ a constant value for each system, in order to better examine $\Delta\rho/\rho_0$'s temperature evolution without the dominant trend of the power law form.

In single-Q Cr, there is no convergence to a scaling form up to 180 K (Fig. 4). For Cr above 200 K, the SDW gap gradually closes, and the carrier density increases; Kohler's scaling based on $H/\rho_0 = nec\omega_c\tau$ is not expected to hold beyond that. Below 200 K, the violation of Kohler's scaling can be understood by returning to Fig. 2b, the plot of linear and quadratic components $\Delta\rho_{lin}$, $\Delta\rho_{quad}$ versus $\rho_0(T)$. If Kohler's scaling is to hold, one would expect $\Delta\rho_{lin}(T,H)$, being proportional to $H$, to be independent of $\rho_0$, while $\Delta\rho_{quad}(T,H)$ evolves $\sim H^2/\rho_0(T)$. Instead, $\Delta\rho_{quad}$ is close to a constant value of $\rho_0$, while $\Delta\rho_{lin}$ is highly sensitive to $\rho_0$, nearly diverging as $H/\rho_0(T)$. This relative relationship of linear and quadratic components is totally opposite to what is expected from Kohler's scaling. The presence of linear MR, with its origin from a sharp curvature, thus breaks Kohler's scaling.

Kohler's scaling in Mo is satisfied over the full field range for $T > 30$ K (Fig. 4). Given Mo's Debye temperature ~400 K, the Kohler scaling from 30 to 90 K is justified with no significant change in the contribution from phonons. Below that temperature, $\Delta\rho/\rho_0$ spread out at high field and differ from the master curve $f(H/\rho_0)$, and from each other. They trace back onto the master curve in the low field limit at all temperatures. This is expected as at zero field $r$ becomes infinitely long in the Onsager-Lifshitz relationship for all carriers regardless of the local curvature, and much longer than the free path length $l$. The temperature Kohler's scaling emerges over the full field range of interest thus sets the upper temperature boundary to the MR anomaly. Hence $T \sim 30$ K is the characteristic temperature for Mo at 14 T where the SdH oscillation disappears (Fig. 3) and Kohler's scaling starts to emerge (Fig. 4b), Importantly, the MR anomaly also disappears at this temperature at 14 T (Fig. 3e). The transport processes that contribute to the MR anomaly are bound between these two scenarios so that carriers do not necessarily complete a cyclotron movement of the quantum orbit, but nevertheless retain its local characteristics and separate from the globally uniform behavior.

In the limit of extremely low field, $\Delta\rho/\rho_0$ approaches the parabolic field dependence $(H/\rho_0)^2$. Nevertheless, the master curve $f(H/\rho_0)$ of Kohler's scaling extends beyond the simple $H^2$ form but has a non-trivial shape that is readily visible in Fig. 4b of $\frac{\Delta\rho}{\rho_0}/(\frac{H}{\rho_0})^{1.8}$ vs. $H/\rho_0$. A non-quadratic form of Kohler's scaling has been documented in the literature [27, 28]. Kohler's scaling is based on a uniform relaxation time $\tau$ for all carriers [24], but the shape of curve $f(H/\rho_0)$ would deviate from the quadratic form $H^2$ when the system moves from low to intermediate field range as carriers can make motions close to a full Fermi surface orbit. The shapes of the Fermi surface thus affect the functional form $f(H/\rho_0)$. While SdH quantum oscillations depend on the field $H$ but not $H/\rho_0$ [8], the general MR is still affected by the cyclotron motion in semiclassical transport theory. Here in the intermediate field range, where Landau level based quantum transport and semiclassical transport theory meet, it is also where the low-$T$ MR anomaly in Mo starts to deviate from the master curve $f(H/\rho_0)$ (Fig. 4b).

**DISCUSSION**

Cr and Mo are two complementary model systems to reveal the origin of the low-$T$ MR anomaly. The SDW in Cr creates an open orbit because of the partially opened SDW gap [18]. As the gap size is stable to 200 K and the open orbit is always present below $T_N$=311 K, the Fermi surfaces in Cr are stable over the temperature range below 100 K where the low-$T$ MR sharply varies (Fig. 1). For current **I** parallel to SDW wavevector **Q** along the *a*-axis in real space, $\rho_{aa}$ is not affected by the open orbit but can rise and not saturate [18, 24]. Conventionally, it would increase as $H^2$, but the power law exponent in Cr around 1.3 indicates the detailed transport process for large MR is determined by the sharp corner within the general conditions of an unsaturated MR.

Either compensation or open orbits can lead to an unsaturated MR, but these conditions alone do not clarify the microscopic origin of a large MR [1]. The stable configurations of compensation or the open orbit with changing temperature do not explain the low-$T$ MR anomaly. Furthermore, the electron-hole compensation in Mo (and W) is insensitive to disorder, but the size of the MR is strongly dependent on disorder. As evidenced by the strong correlation between $\rho_0$ and *RRR* values of all metals listed in Fig. 1, increasing disorder would reduce the relaxation time $\tau$ (and length $l$) and make the sharp corner mechanism of enhancing MR [7, 8, 20, 21] less effective. In addition, increasing disorder can alter the lattice coherence length, which makes the Fermi surface less well defined and smears out its sharp features. Both scenarios have disorder affecting the sharp corner feature. For systems approaching the dirty limit, deformed Fermi surfaces with smeared sharp features can be significant. As free electrons of parabolic band dispersion do not support MR [8], non-vanishing MR necessarily involves Fermi surface features with non-parabolic dispersion such as the sharp corners [29]. As systems approach the clean limit, the first effect likely would dominate as both $\tau$ and MR can diverge with a growing lattice coherence length and increasing *RRR* values. For the intrinsic sharpness of Fermi surface features, there might not be significant improvement beyond a certain limit of decreasing disorder, except in the scenario of two superposed lattices in DW-based systems. For example, the sharpest feature of the Fermi surface is estimated ~0.004 Å$^{-1}$ in our Mo specimen of *RRR*~2900, while the sharpest feature is about ~3 10$^{-5}$ Å$^{-1}$ in our Cr specimen of *RRR*~70 because of the superposed SDW state and the lattice. Sharp curvatures of the Fermi surface, under unsaturated MR conditions due to either open orbits or compensation, thus provide the microscopic mechanism of an enhanced MR magnitude that is characterized by the low-$T$ MR anomaly.

Our proposed universal origin for the low-$T$ MR anomaly is characterized by the competition between the carriers' relaxation lengths $l$ and the sharp curvature's radius. This discussion so far has been based on geometrical arguments, with a simple assumption that the carrier's relaxation length $l$ is isotropic and independent of the carrier's location on the Fermi surface orbit. This perspective of a local $\omega\tau$ is equivalent to the carrier's local mobility $\mu$ being expressed as $\tau = H\mu$, where $\mu = e\tau/m^*$ with $m^*$ the carrier's mass. While higher mobility $\mu$ can be due to carriers of lighter masses, those light carriers are closely correlated with small dHvA orbits [8, 30] and SdH pathways involving quantum interference [19]. As $m^* = \frac{\hbar^2}{2\pi}(\partial A/\partial \varepsilon)_k$, a small orbit of limited area $A$ and thus a small $r_k$ is often strongly correlated with a small local cyclotron mass. Hence small orbits, or simply a sharp arc with a large curvature, can naturally have carriers of higher mobility than the other segments of the Fermi surface.

The temperature dependence of the low-$T$ MR anomaly places the phenomenon between quantum oscillations and semi-classical transport theory. Local and sharp Fermi surface curvature represents a non-dissipative mechanism that increases the MR. In the low-$T$ limit, magnetoresistance rises under high field because carriers perform many cycles of the quantum orbit motion before being scattered by disorder. The adverse effect of disorder in destroying MR is similarly observed in many forms of quantum transport across different magnetic field regimes. It was pointed out by Abrikosov in Ref. [14] that, in the extreme quantum limit, disorder would cause increased scattering/hopping of carriers in quantized cyclotron motion around one field line to another; this effectively increases conductivity and reduces MR. This argument also applies to the disorder effect on the zero longitudinal conductance state in the integer quantum Hall effect and can be extended to carriers of cyclotron motion even if they are far away from the extreme quantum limit, such as those in our measurements.

**Methods**

**Cr and Mo single crystals.** Our single-crystal Cr transport sample was prepared from a wafer of ~2 mm thick and 10 mm diameter with a surface normal of (1, 0, 0), previously purchased from Alfa Aesar (99.996+%, Lot #13547). After diamond wheel saw cutting, alumina suspension polishing, thermal annealing, and acid etching [18], the final electrical transport sample had a length of ~5 mm, with a rectangular cross-section of 1.84 × 0.42 mm$^2$. This sample was studied previously in Ref. [18] where more information about the preparation of both the sample and a single-Q state can be found. The residual resistivity ratio ($RRR$) with **I** ∥ **Q** is $\rho(T = 350K)/\rho(T = 2K) \sim 70$.

A wafer of Mo single crystal was purchased from Goodfellow (99.999%, MO002129) with a 12 mm diameter and a (1, 0, 0) surface normal. Several matchstick shaped samples were prepared by electron-discharge machining (EDM) and diamond wire saw cutting, with the single crystal's orientation determined by lab x-ray Laue diffraction. The sample was further fine polished using 50 nm alumina suspension (MicroPolish, Buehler Ltd.) to an optical finish and etched in a hot 10% HCl bath to remove potential surface damage. The final single crystal sample measured 12.3× 0.90×0.38 mm$^3$, with every surface normal a cubic axis (1, 0, 0) type.

A matchstick-shaped W single crystal was inherited from the Institute for the Study of Metals at the University of Chicago. The crystal orientation was x-ray aligned with (1, 0, 0) type of cubic axes along length, width, and depth directions, and was cut by wire saw to a shape of 0.62×0.74×21.0 mm³. The sample might have residual amount of amorphous W that can superconduct [16], which shows up in the MR behavior at low field and affects the precise determination of $\rho_0(T, H = 0)$ below 25 K. On the other hand, $\rho_0(T, H = 14T)$ (Fig. 1) is not affected by such superconducting threads.

**Electrical measurements.** Each single crystal sample was anchored to the surface of an 8-pin DIP connector using GE 7031 varnish. The electrical connection was made by gold wires (25μm diameter) and silver epoxy (EPO-TEK H20E-PFC, Epoxy Technology). Samples were placed inside a 14-T Physical Property Measurement System (PPMS DynaCool-14, Quantum Design Inc.) from 1.65 to 355 K and magnetoresistance were measured using an AC resistance bridge and a preamplifier (LS372 and 3708, Lake Shore Cryotronics, Inc.). To achieve the best stress-free condition and, in turn, the best residual resistivity at low temperature, we always cool down samples slowly, typically below 1 K/min. At each field, multiple measurements were repeated to achieve a highly precise value of $\rho$.

For our matchstick samples, the antisymmetric Hall resistance is negligible between the transverse MR leads. However, there still exists a finite remanent field in PPMS's superconducting magnet, caused by field sweeps over a 14T range. To eliminate this distortion, we performed all MR measurements from 14 T to about -0.5T, and a shift of the field zero on the order of tens of Oe was applied by making the MR symmetric at the low field. This correction was not required for the SdH measurements of Mo as the oscillations were measured at fields above 6 T. For the MR measured in the low-field limit in Figs. 2e and 3f, the remanent field is removed by many rounds of field oscillations before the MR measurement was performed.

**Figure captions:**
**Fig. 1.** Low-temperature magnetoresistance anomaly in elemental metals. Magnetoresistance as a function of temperature measured for (a) Cr, (b) Mo, and (c) W, with all field and current directions aligned along the cubic axes. (d) At low temperature, $\rho(T)$ of both Cr and Mo demonstrate a disorder-dominated $T^3$ power-law dependence (purple-colored points). For the Mo specimen with an improved *RRR* value, the temperature range of the $T^3$ power-law dependence is much reduced. (e, f) The low-temperature MR anomaly is significantly reduced in Mo and W polycrystal specimens of lowered *RRR* values, indicating disorder's adverse effect.

**Fig. 2.** Features of the low-*T* MR anomaly in Cr. (a) Over a fixed low field range, $\rho(H)$ changes its functional form from linear to parabolic with increasing temperature. (b) Linear and quadratic components of $\rho(H)$ are separated as $\Delta\rho(H) = \rho(H) - \rho_0 = \Delta\rho_{lin}(H) + \Delta\rho_{quad}(H) = AH + BH^2$ for data over the fixed field range in panel (a), and plotted against $\rho_0(H = 0)$. (c) SdH oscillations of SDW Cr in the single-Q state at 1.65, 50, and 90 K, adapted from the full data set in Ref. [18]. A low frequency 36 T SdH oscillation is observed down to 1.5 T at 1.65 K and also survives to above 90 K. (d) All three galvanomagnetic behaviors, namely linear MR $\Delta\rho_{lin}(H)$, SdH oscillation amplitude of the 36T orbit, and MR at 14T $\Delta\rho(H = 14T)/\rho(H = 0)$, are plotted as a function of temperature. Their coincidence identifies the 36 T SdH orbit as responsible for the MR anomaly. (e) Magnetoresistance at low field and 1.65 K. The data around zero field (red points) are fit to a pure parabolic form (red solid line), which marks a critical field $H_c \sim 135$ Oe. This fitting leads to an estimation of a sharp curvature of $10^{-4}$-$10^{-5}$ Å$^{-1}$ on the 36 T orbit that is responsible for the low-$T$ MR anomaly.

**Fig. 3.** Galvanomagnetic characteristics of Mo. (a) SdH oscillations of Mo are compared between 1.65 K and 9 K. Red triangles indicate quantum oscillation frequences that are also observed in dHvA techniques, while white triangles indicate higher harmonics during the SdH measurement. (b) The 1688 T oscillation is the only observed SdH frequency that has no dHvA correspondence and is not a harmonic, and remains observable at 28 K, indicating its quantum interference nature. (c) Raw SdH oscillations at 18 and 24 K demonstrate that the 518 T mode is still observable at 18 K but not at 24 K. (d) Integrated intensities of three SdH orbits that survive above 9 K. The high temperature part of each temperature dependence (solid points) can be fit with the Lifshitz-Kosevich formula. (e) Temperature dependences of the MR anomaly $\Delta\rho(H = 14T)/\rho(H = 0)$, compared to scaled amplitudes of the 518 and 1688 T SdH oscillations. The SdH oscillations set a lower boundary of $\Delta\rho(14T)/\rho$'s temperature scale. (f) Low-field MR of Mo at the base temperature. The MR data between ±420 Oe (red points) can be fit to a pure parabola (red curve). The field scale and $\rho_0$ provide an estimation of the sharpest curvature on the 518 T orbit.

**Fig. 4.** Kohler's scaling plots. Magnetoresistance of (a) Cr and (b) Mo are displayed in the log-log scale to examine both forms of the power law and Kohler's scaling. "Residual" plots of $\frac{\Delta\rho}{\rho_0}/(\frac{H}{\rho_0})^\alpha$ vs. $H/\rho_0$ are also provided to illustrate fine details of the scaling function. The temperature range where Kohler's scaling emerges over the full field range of interest sets the upper temperature boundary to the MR anomaly.

**Acknowledgments:** Y. F. acknowledges the support from Okinawa Institute of Science and Technology Graduate University with subsidy funding from the Cabinet Office, Government of Japan. Y.W. acknowledges the startup support from the University of Tennessee, Knoxville. T.F.R. acknowledges support from the US Department of Energy Basic Energy Science Award No. DE-SC0014866. P.B.L. is funded by the Gordon and Betty Moore Foundation, Grant GBMF12763. H.C. acknowledges support from US NSF CAREER grant DMR-1945023.

**Author Contributions:** Y. F. designed research. All authors performed research and analyzed data. Y. F. prepared the manuscript, and all authors commented.

**Competing Interest Statement:** The authors declare no competing interests.

**Data availability:** All study data are included in the article and/or supplementary materials.

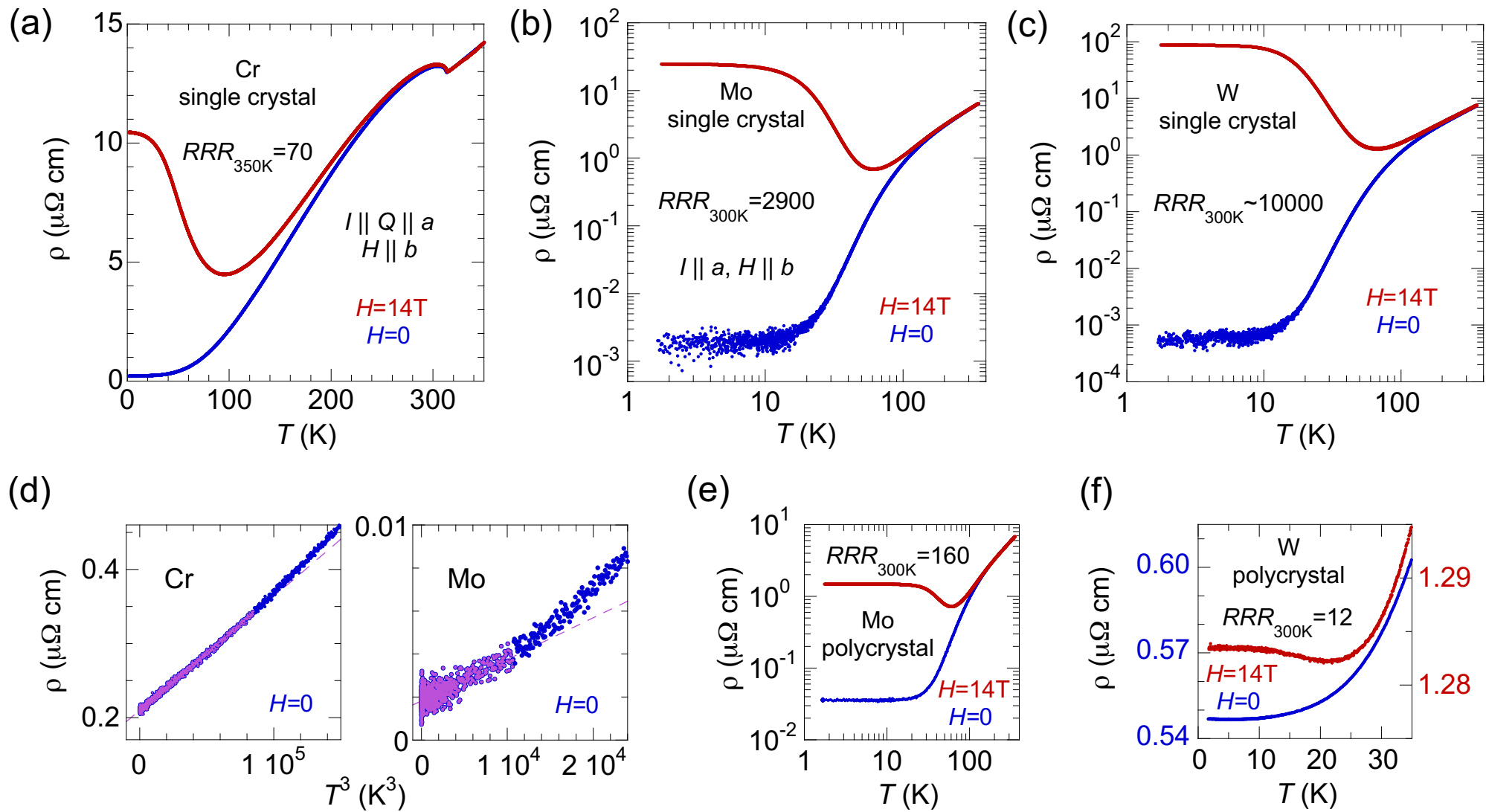

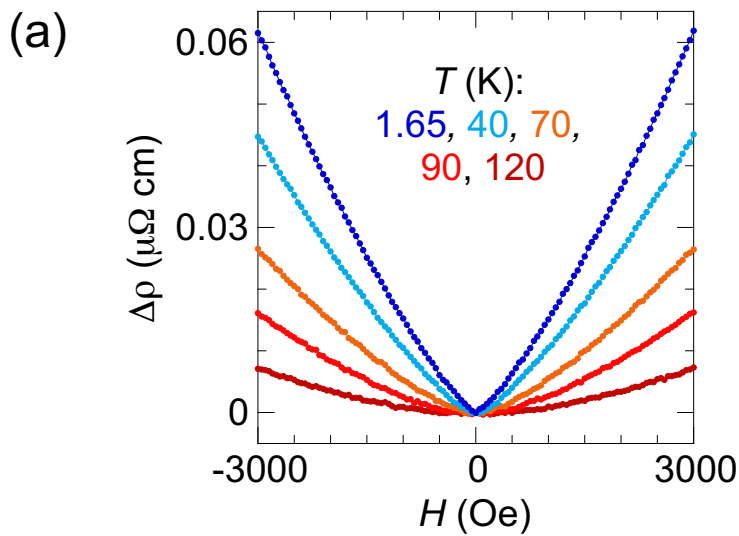
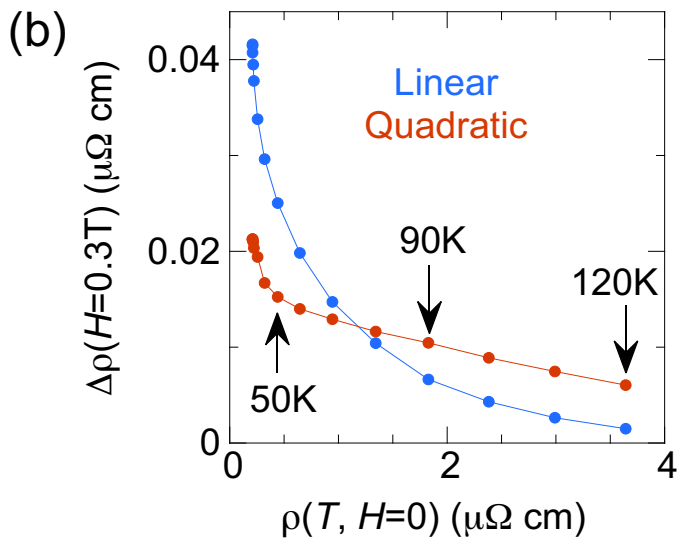
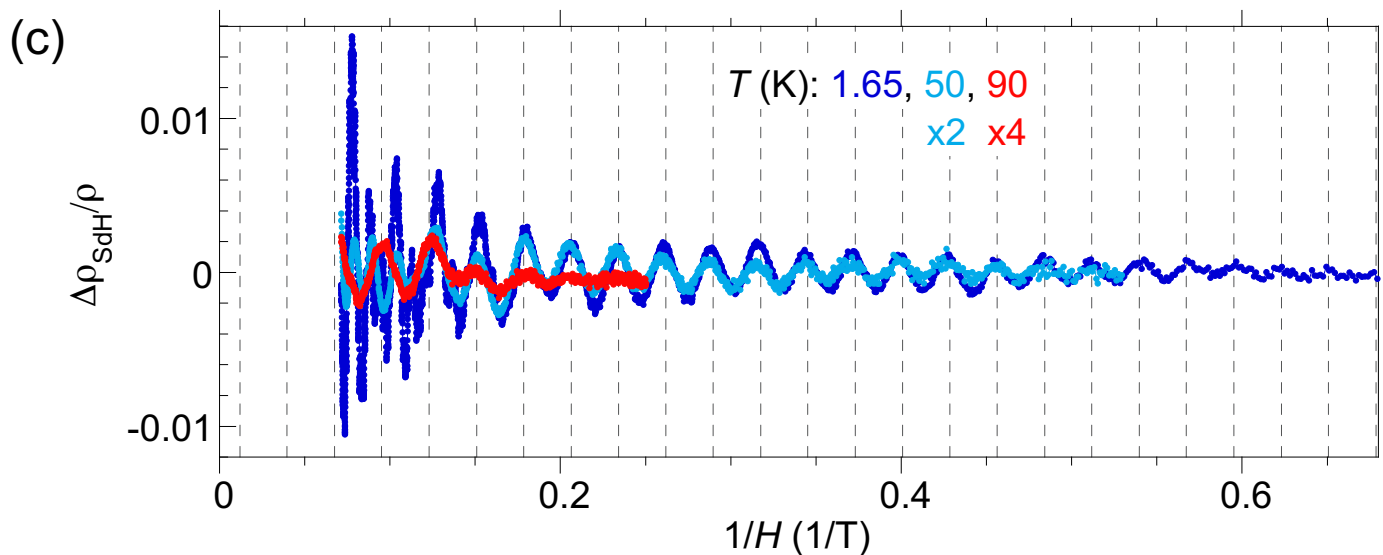
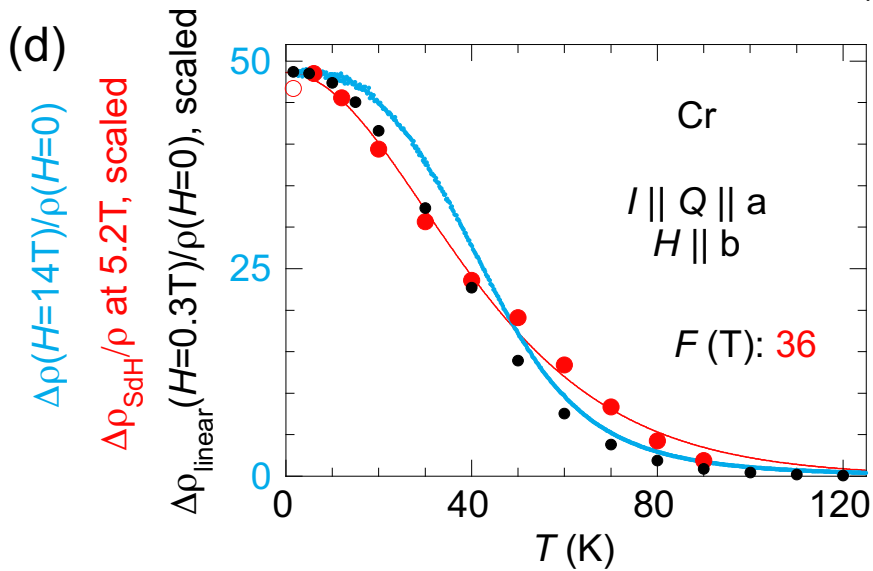
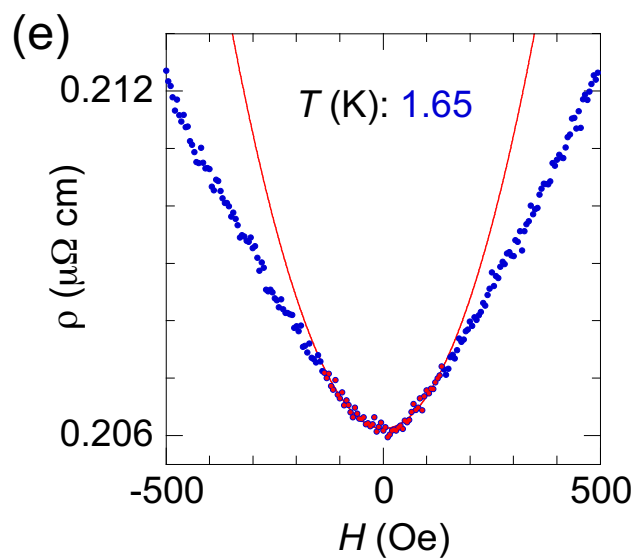

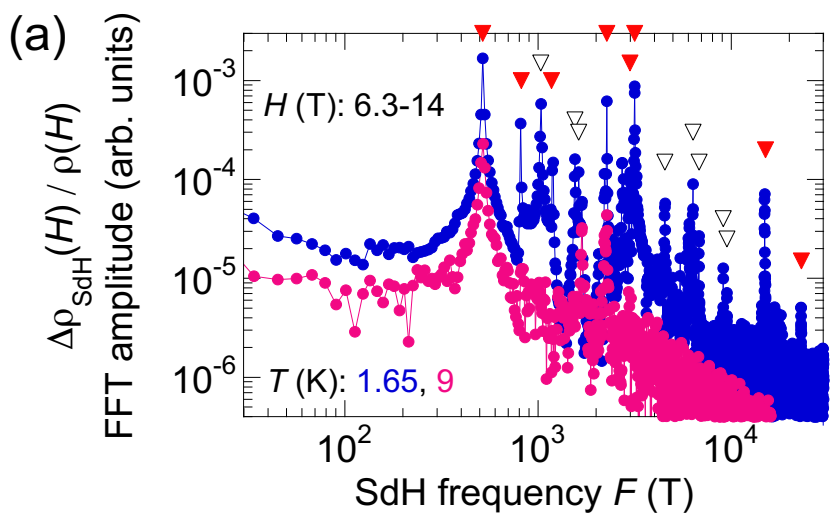
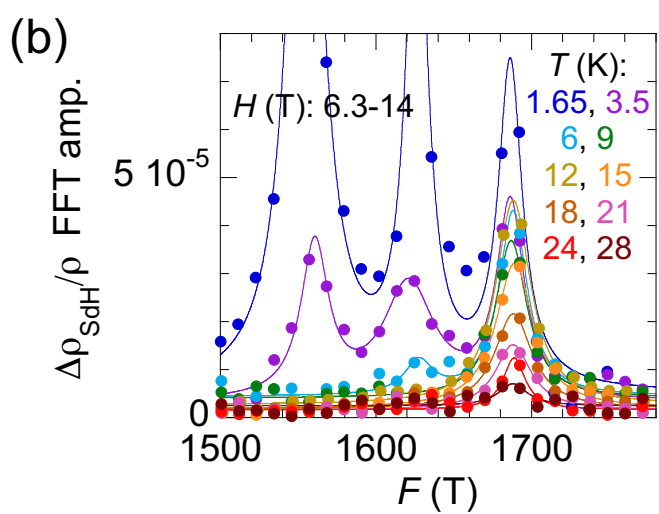
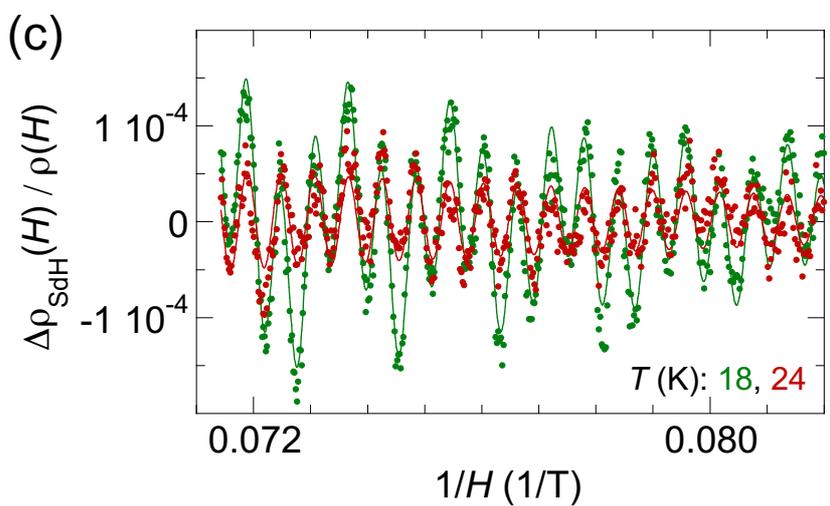
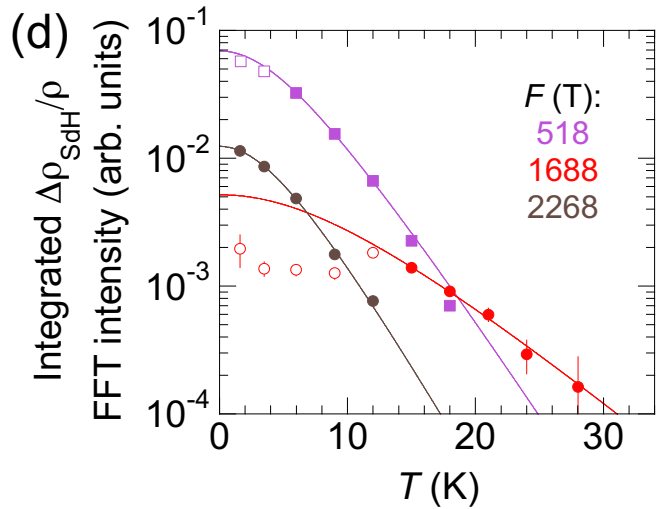
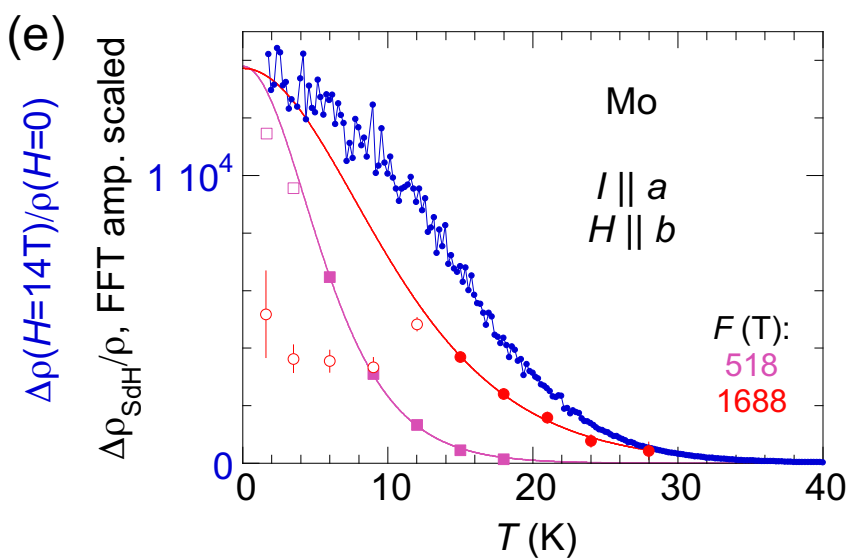
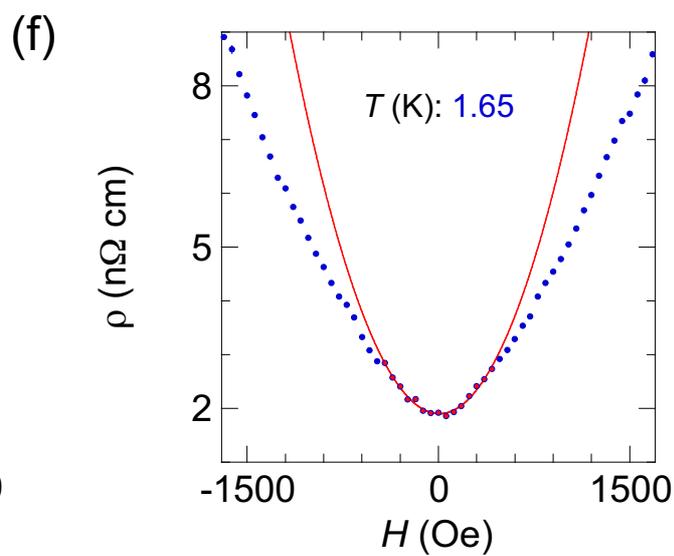

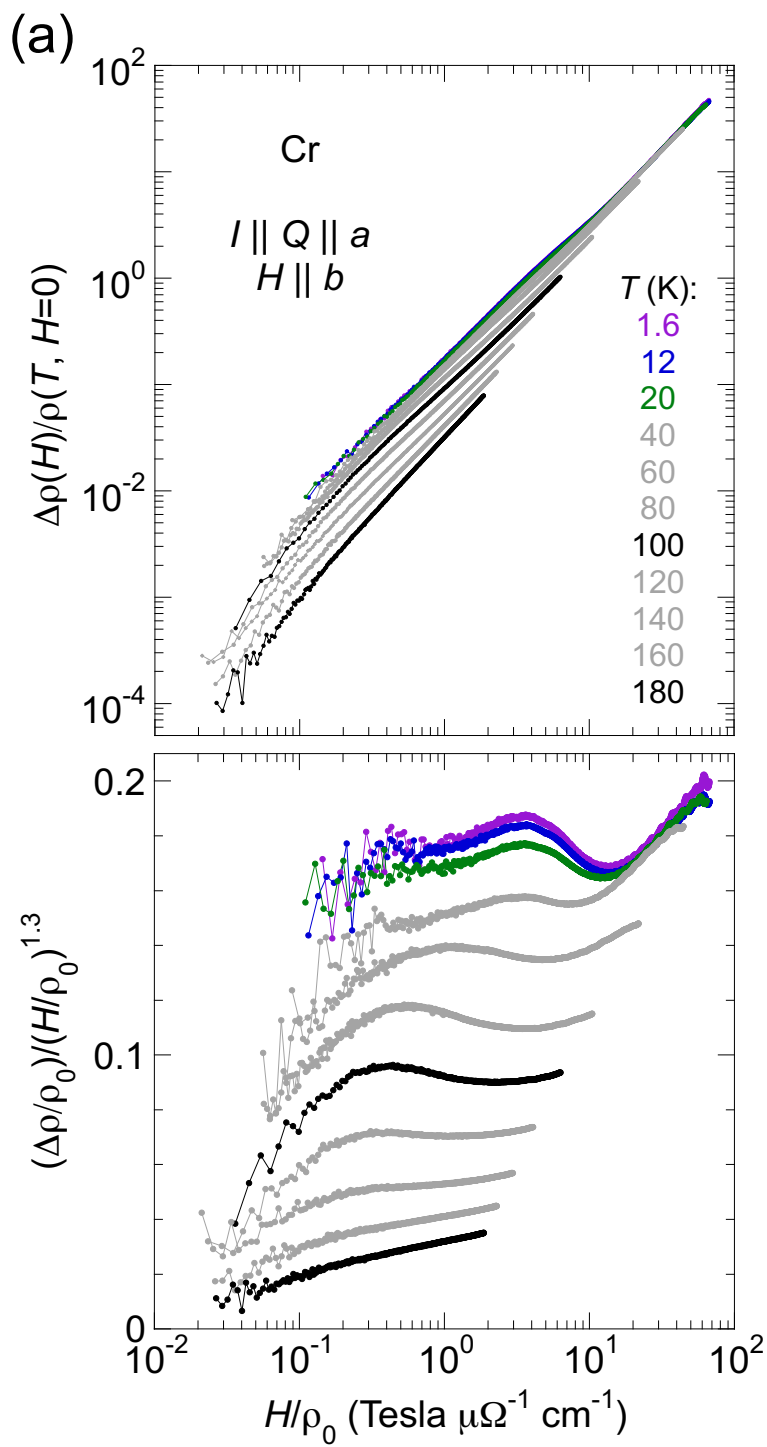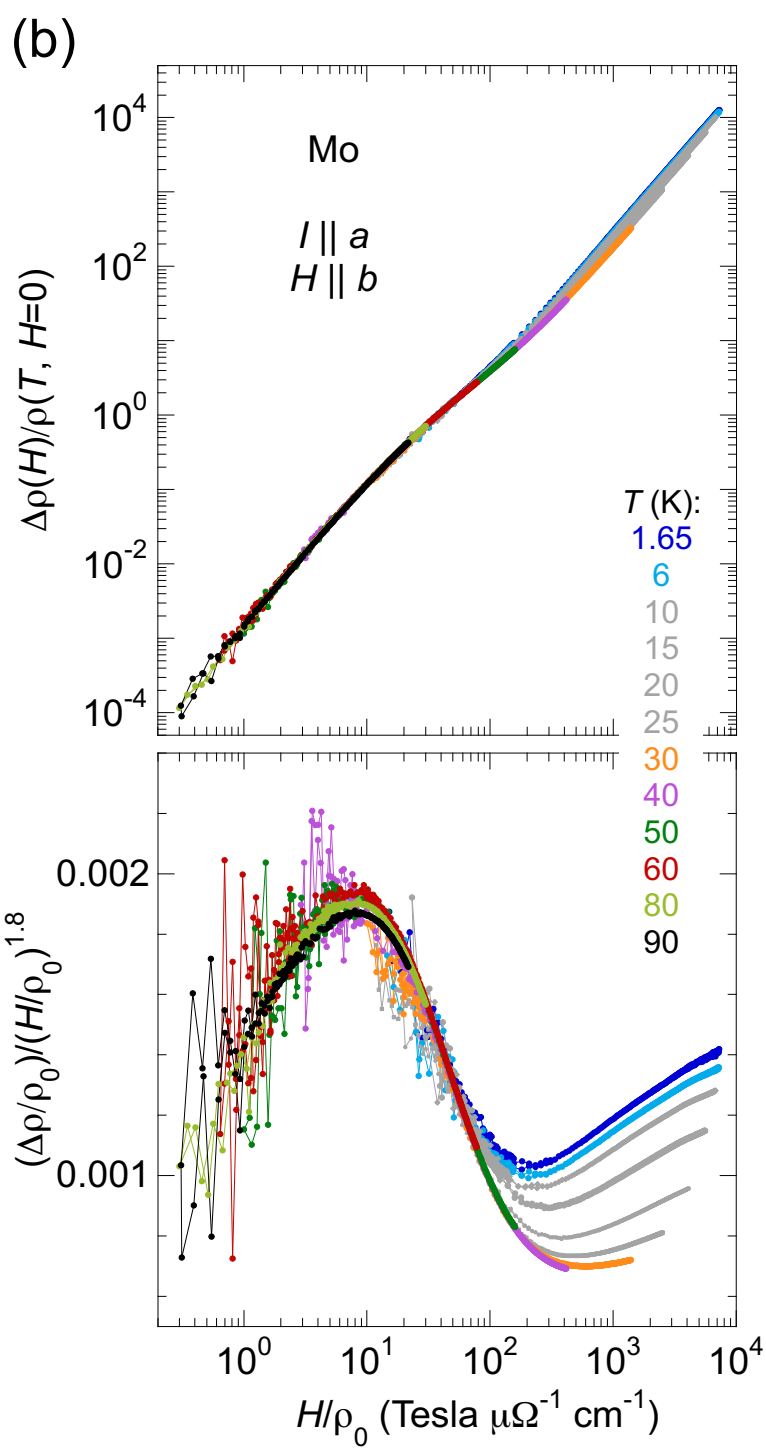